\documentclass[prl,twocolumn,showpacs,preprintnumbers,a4paper,superscriptaddress]{revtex4-1}
\usepackage{amsmath}
\usepackage{epsf}
\usepackage{mathtools}
\usepackage{amsfonts}
\usepackage{amssymb}
\usepackage{color}
\usepackage{graphicx}
\usepackage{hyperref}

\def\m@thcombine#1#2{
  \setbox0=\hbox{$#1$}
  \setbox1=\hbox{$#2$}
  \ifdim\wd0>\wd1
    \setbox0=\hbox to\wd1{\hss\box0\hss}
  \else
    \setbox1=\hbox to\wd0{\hss\box1\hss}
  \fi
  \mathop{\vcenter{
    \offinterlineskip\box0\box1}}}
\def\gesim{\m@thcombine>\sim}

\begin{document}

\title{Unbinding of giant vortices in states of competing order}
\author{ J. M. \surname{Fellows} }
\affiliation{School of Physics and Astronomy, University of Birmingham, Birmingham B15
2TT, United Kingdom}
\author{ S. T. \surname{Carr} }
\affiliation{Institut f\"ur Theorie der Kondensierten Materie and DFG Center for
Functional Nanostructures, Karlsruher Institut f\"ur Technologie, 76128
Karlsruhe, Germany}
\author{ C. A. \surname{Hooley} }
\affiliation{Scottish Universities Physics Alliance, School of Physics and Astronomy,
University of St Andrews, North Haugh, St Andrews, Fife KY16 9SS, United
Kingdom}
\author{J. \surname{Schmalian}}
\affiliation{Institut f\"ur Theorie der Kondensierten Materie and DFG Center for
Functional Nanostructures, Karlsruher Institut f\"ur Technologie, 76128
Karlsruhe, Germany}

\begin{abstract}
We consider a two-dimensional system with two order parameters, one with O(2) symmetry and one with O($M$), near a point in parameter space where they couple to become a single O($2+M$) order.  While the O(2) sector supports vortex excitations, these vortices must somehow disappear as the high symmetry point is approached.  We develop a variational argument which shows that the size of the vortex cores diverges as $1/\sqrt{\Delta}$ and the Berezinskii-Kosterlitz-Thouless transition temperature of the O(2) order vanishes as $1/\ln(1/\Delta)$, where $\Delta$ denotes the distance from the high-symmetry point. Our physical picture is confirmed by a renormalization group analysis which gives further logarithmic corrections, and demonstrates full symmetry restoration within the cores.
\end{abstract}

\pacs{64.60.Bd, 05.10.Cc, 05.70.Jk}
\maketitle

Competition between different states of order is a hallmark for a large
class of strongly correlated systems. Some (quasi-)two dimensional examples are competing magnetism and
superconductivity in cuprates \cite{cuprates}, organics \cite{organics} and
the recently discussed thin-film heavy-electron systems \cite{heavyelectron};
competing superfluid and crystalline order that may occur in bosonic systems
on triangular lattices \cite{he3} or cold atomic gases \cite{fellowscarr2011,lecheminant};
or charge density wave order that competes
with superconductivity in Sr$_{14-x}$Ca$_{x}$Cu$_{24}$O$_{41}$ \cite{SrCuO}. 
In all these cases, at least one of
the two competing order parameters has XY, i.e.\ O(2)
symmetry, while the other may in general be O($M$). The cases $M=1$, $2$, and $3$ correspond to the other order parameter being of Ising, XY, or Heisenberg type, respectively. The $M=1$ case
describes easy-plane magnetism \cite{nelsonpelcovits,Pelissetto-Vicari-2007}; the case $M=2$ relates to
supersolid phases in cold-atom
systems \cite{fellowscarr2011} and to competing density-wave
and superconducting order in layered materials \cite{Carr-Tsvelik-2002,rozhkovmillis2002,Jaefari-Lal-Fradkin-2010}; models with 
$M=3$ have been considered in the context of
high-temperature superconductivity \cite{Demler-Hanke-Zhang}. Both order
parameters interact and the symmetry of the coupled problem is O($M$) $\times$ O(2).  However, at a certain fine-tuned point in phase space one may expect the symmetry
to be enhanced, from O($M$) $\times$ O(2) to O($N$) with $N=M+2$.  This is not the most general scenario for competition between two order parameters, but it has been conjectured to occur in many different microscopic models, including all of the cases mentioned above \cite{fellowscarr2011,lecheminant,SrCuO,nelsonpelcovits,Pelissetto-Vicari-2007,Carr-Tsvelik-2002,rozhkovmillis2002,Jaefari-Lal-Fradkin-2010,Demler-Hanke-Zhang}.

This symmetry enhancement acquires a particularly interesting
aspect in layered or two-dimensional systems, where long range order is
absent for continuous symmetries due to the
Hohenberg-Mermin-Wagner theorem \cite{Hohenberg-Mermin-Wagner}.
However, the O(2) sector supports non-trivial
topological configurations, i.e.\ vortices.  The
unbinding of vortex-antivortex pairs converts an
algebraically ordered superfluid or crystal to a disordered normal fluid.  This
Berezinskii-Kosterlitz-Thouless (BKT) transition \cite{b,kt,k} occurs at a non-zero temperature 
$T_{\mathrm{BKT}}$.

\begin{figure}[tbp]
\begin{center}
\includegraphics[width=2.8in,clip=true]{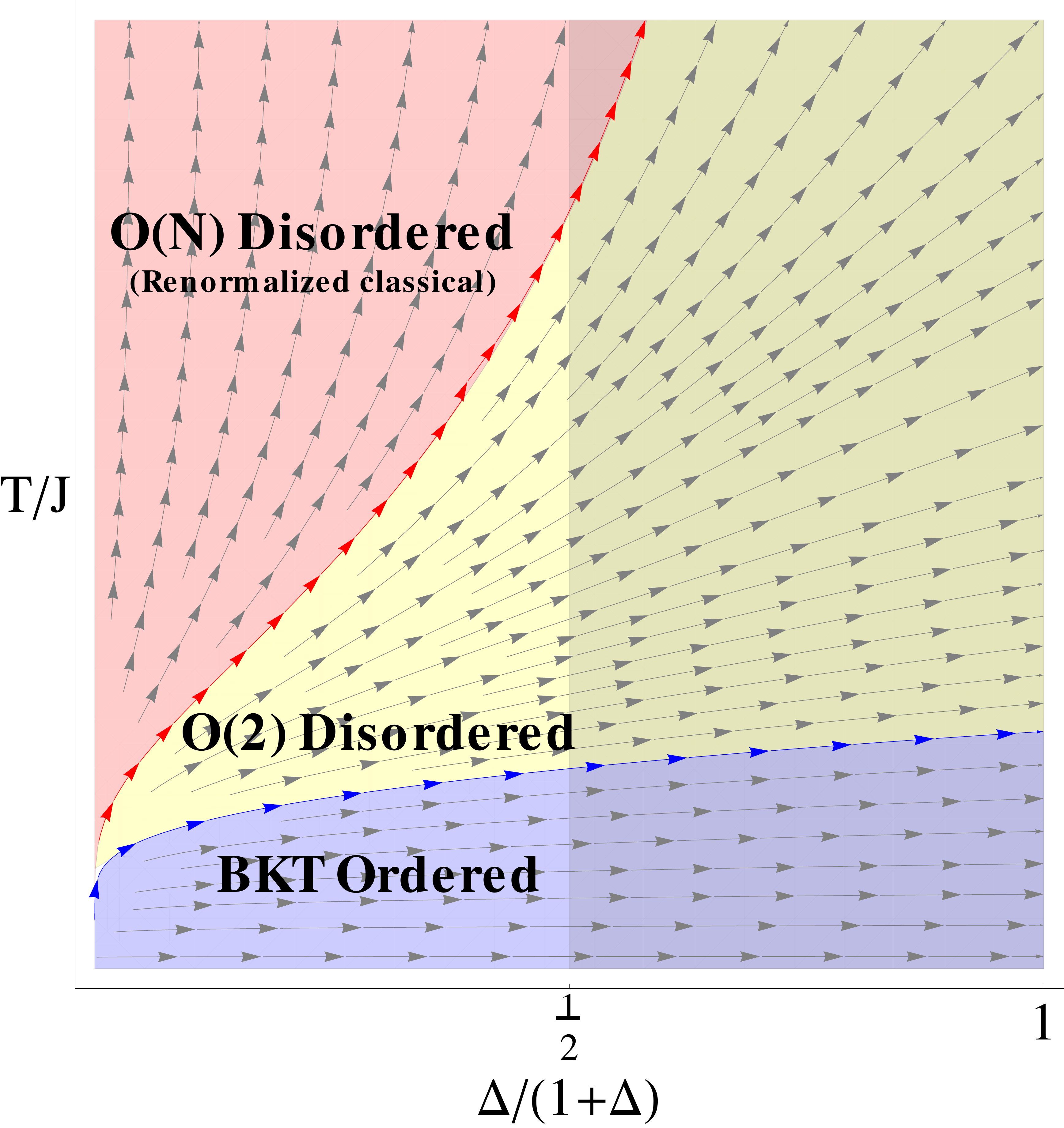}\vspace{-0.3cm}
\end{center}
\caption{[Color online] Schematic phase diagram of the model (\protect\ref{nlsm1}) for $\Delta>0$  near $\Delta=0$.  The thick blue line indicates $T_{\rm BKT}$ as a function of $\Delta$.  Below this line the model shows power-law correlations; above the line it is XY-disordered.  The red line shows the crossover temperature between a disordered state of O(2) character (proliferation of vortices) and one of O($N$) character (spin waves). The crossover line occurs where the size of the vortex core becomes of the same order as the average distance between vortices.  The arrows show the RG flow (\ref{RG-flow-equations}) for the specific case $M=2$ (see text). The shadowed region, $\Delta\gtrsim1$, indicates where spin wave renormalization must be complemented by vortex considerations.}
\label{Fig1}
\end{figure}

Suppose the
fine tuning to a high-symmetry point in the phase diagram is achieved by
varying a dimensionless parameter $\Delta>0$ towards $\Delta =0$ which
corresponds to the O($N$) symmetry point.  In each realization of this model, the experimental handle corresponding to our parameter $\Delta$ is different -- for example in the context of cuprates/organics it would correspond to doping/pressure \cite{Demler-Hanke-Zhang} while for cold dipolar bosons it may be controlled via the angle of a polarizing field \cite{fellowscarr2011}.
However, no matter which particular microscopic realization of this model is chosen, if $N>2$ then $T_{\mathrm{BKT}}$ must
vanish for $\Delta \rightarrow 0$. Indeed, combining spin-wave based
renormalization group calculations with crossover arguments one can estimate
that $T_{\mathrm{BKT}}$ vanishes as $1/ \log(1/\Delta)$
as $\Delta $ vanishes -- see Fig.~\ref{Fig1}.  This was first derived for the case $M=1$ in \cite{nelsonpelcovits}; the present work extends this result to generic values of $M$.
 There are, however, a number of
nontrivial aspects that emerge from this picture.  The BKT transition must
vanish because of the dominance of spin wave excitations of the high
symmetry model.  On the other hand, spin waves do not usually interfere with vortices:\ in
the O(2) model, spin waves do not lead to a renormalization
of the stiffness. So how do vortices become spin waves? 

In this paper we investigate the fate of vortices of XY-order
parameters, and of the BKT transition that they mediate, as the high-symmetry point is approached. Combining
variational arguments and a renormalization group (RG) analysis, we study the
crossover and transition temperatures and show that for small $\Delta$, the size of the
vortex core diverges as
\begin{equation}
\xi _{0}\simeq a\frac{\ln ^{1/M}\left( 1/\Delta \right) }{\sqrt{\Delta }},
\label{core at BKT}
\end{equation}
where $a$ is the core size of a single O(2)
order parameter (i.e.\ for $\Delta \gesim 1$). This is a consequence of the
emergence of competing order and  O($N$) spin waves inside the vortex core. Thus, the
enhanced symmetry becomes visible not only at high temperatures, where $\Delta$ may be neglected, but also at low temperatures, via the size of the
vortex core near (and below) the unbinding transition.

In accordance with the usual ideas of universality, we consider the long wavelength action with the appropriate symmetry, which is a perturbed non-linear sigma model:
\begin{equation}
S=\frac{J}{2T}\int d^{2}x\left[ \left( \nabla \mathbf{n}\right) ^{2}+\frac{\Delta }{a^{2}}\mathbf{n}^{T}D\mathbf{n}\right]  \label{nlsm1}.
\end{equation}
Here $\mathbf{n=}\left( \mathbf{s,m}\right) $ is an $N$-component vector subject to the unit-length constraint $\mathbf{n}^{2}=1$. 
The vector $\mathbf{s}$ has two components while $\mathbf{m}$ has the remaining $M$ components; these two vectors corresponding to the two competing order parameters of the original theory.
The matrix $D$ is given by
\begin{equation}
D=
\begin{pmatrix}
0 & 0 & 0 \\ 
0 & 0 & 0 \\ 
0 & 0 & \widehat{1}_{M\times M}
\end{pmatrix}.  \label{anisotropymatrix}
\end{equation}
The model at $\Delta =0$ has full O($N$) symmetry, whereas for $\Delta \neq 0
$ this symmetry is broken to O($M$) $\times$ O(2) by giving a mass to the ${\bf m}$ sector of the theory. The action contains a
reference length scale, $a$ (of the
order of the crystal lattice spacing), and a reference energy scale, $J$. The exact meaning of $J$
depends on the microscopic model from which (\ref{nlsm1}) is derived; but
typically $J$ corresponds to a bandwidth of the unperturbed microscopic
model, possibly reduced by quantum fluctuations or geometric frustration.
We assume we are in
a regime where the ground state is fully ordered so that any possible quantum dynamics
beyond (\ref{nlsm1}) may be safely neglected.

Let us start by considering the situation far from the O($N$) point, i.e.\
for sufficiently large values of $\Delta$. In this case the vector $\mathbf{n}$ is effectively constrained to lie in the easy plane of the $\Delta$-term, and thus becomes a two-component one. Hence the standard BKT argument \cite{b,kt}
applies here, and the relevant excitations to consider are vortices, which are described by a configuration 
\begin{equation}
\mathbf{s}=(n_{1},n_{2})=(\cos \theta ,\sin \theta ),  \label{vortexvector}
\end{equation}
where $(r,\theta)$ are plane polar co-ordinates with $r=0$ at the center of the vortex.
The action for a single such vortex is
\begin{equation}
S_{\mathrm{v}}=\frac{\pi }{T}\ln \left( \frac{L}{a}\right) +S_{\mathrm{core}},  \label{vortexaction}
\end{equation}
where $L$ is the linear size of the system, $a$ is the lattice spacing, $S_{\mathrm{core}}\left(T\right) \sim \pi /2T$ is the action of the
(lattice-scale) core of the vortex, and we work henceforth in units where $J=1$. In (\ref{vortexaction}) we were obliged to cut off the divergent
energy near the center of the vortex at the lattice scale $a$, because the
model contained no other length scale. 
As $\Delta$ is reduced, however, a
new mechanism of removing this divergence becomes available:\ the vector $\mathbf{n}$ 
can simply be rotated away from the easy plane \cite{bigcores}. This costs an
energy proportional to $\Delta$, but is worth doing near the center of the
vortex where the vortex action would otherwise be very high. Let us, then,
consider a configuration described by a new length-scale $\xi$:\ for $r>\xi$,
 it is identical to the above-described vortex, while for $r<\xi$, $\mathbf{n}$ has components perpendicular to the easy plane. The action of
such a configuration may be estimated as
\begin{align}
S & \approx S_{\mathrm{core}}+S_{\mathrm{vortex}}  \notag \\
& =\frac{\pi }{T}\frac{\Delta }{2a^{2}}\xi ^{2}+\frac{\pi }{T}\ln \left( 
\frac{L}{\xi }\right) ;  \label{modvortexaction}
\end{align}
minimizing this with respect to $\xi$ determines the optimum core
size,
\begin{equation}
\xi _{0}=\frac{a}{\sqrt{\Delta }}.  \label{xi0}
\end{equation}
This has the interesting consequence that while the core size increases and
the total vortex action decreases with decreasing $\Delta$, the core action
remains the same:\ $S_{\mathrm{core}}=\pi /2T$, as in the original BKT case.

The finite-temperature transition from the quasi-long-range-ordered to the
disordered state occurs via a proliferation of unbound vortices and
anti-vortices. To describe this one considers the  renormalization group flow equations \cite{kt}
\begin{equation}
\frac{dT^{-1}(\ell)}{d\ell} = 4\pi^3 y(\ell)^2,\;\; \frac{dy(\ell)}{d\ell} = \left(2-\pi T^{-1}(\ell) \right)y(\ell),\label{BKT-RG}
\end{equation}
where $T(\ell)$ and $y(\ell)$ are the renormalized temperature and vortex fugacity at length scale $\xi=ae^\ell$.
These flow equations have a separatrix along the line $\pi/2T=1+2\pi y$, meaning that the
 transition temperature satisfies the following
equation:
\begin{equation}
T_{\mathrm{BKT}}=\frac{\pi }{2+4\pi y \left( T_{\mathrm{BKT}}\right) }.
\label{tbkt}
\end{equation}
In the regular single scale BKT transition, the bare fugacity is given by $y(T,\ell=0)=e^{-S_{\mathrm{core}}(T)}$ \cite{kt}, which leads to $T_{\mathrm{BKT}}\approx \pi/2$ as the fugacity gives only a small correction to this value.  As we have seen above, $S_{\mathrm{core}}$ does not depend on $\Delta$, so we must ask the question: how is the BKT transition modified in the presence of another length scale $\xi_0$?  This question is much more general than the model in this paper, and has recently arisen in different situations \cite{orth,she}.

The RG flow \eqref{BKT-RG} should now start at the length scale $\xi_0=ae^{\ell_0}$ and not at the microscopic scale $a$;  so for the initial values we need to know the renormalized temperature and fugacity at this length scale.  Studying the problem just from the point of view of vortices (i.e. ignoring spin waves), the large core derived above is inert, and thus $T(\ell_0)=T(\ell=0)$.  However even for the inert core, the fugacity has a naive scaling dimension and thus flows according to 
\begin{equation}
dy/d\ell=2y;\label{simple-RG}
\end{equation}
integrating this equation gives $y(\ell_0) = (\xi_0/a)^2 y(\ell=0)$.
This enhancement of the fugacity can be understood physically by realizing that while the vortices live at a length scale $\xi_0$, the entropy comes from enumerating the possible positions for the center of the vortex which involves the lattice scale $a$.
The equation for $T_{\mathrm{BKT}}$ now becomes
\begin{equation}
T_{\mathrm{BKT}}=\frac{\pi }{2+4\pi \left( \frac{\xi _{0}}{a}\right)
^{2}e^{-\pi /2T_{\mathrm{BKT}}}}.  \label{tbktmod}
\end{equation}

Substituting in the form (\ref{xi0}) for the optimum core size and defining
a new variable $x=\pi /T_{\mathrm{BKT}}$ we obtain $\Delta =4\pi
e^{-x/2}/\left( x-2\right) $. We are interested in the solution of this
equation as $\Delta \rightarrow 0$, in which case $e^{-x/2}/\left( x-2\right) $
must also tend to zero, i.e.\ $x\rightarrow \infty $. 
Keeping only leading order terms, we see that $x\sim \ln \left( 1/\Delta \right)$, which gives
\begin{equation}
T_{\mathrm{BKT}}\sim \frac{1}{\ln \left( 1/\Delta \right) }.
\end{equation}%
Thus we see that this simple argument gives a BKT transition temperature
that vanishes as $\Delta \rightarrow 0$, as expected on symmetry grounds. 
Furthermore, it defines the length scale $\xi _{0}$:\ below this length
scale the physics of the system becomes sensitive to the proximity to an
enhanced-symmetry point; above it, the physics is essentially that of the
large-$\Delta $, O(2) system.

The analysis above is from the point of view of vortices; it doesn't include spin-waves.
 In particular, we should take into account that the full O($N$) dynamics may still be intact inside the vortex core. To analyze
this issue  we turn to the renormalization group treatment of spin-wave
excitations. The renormalization flow equations of O($N$) nonlinear sigma
models with symmetry broken by giving $M$ of the $N$ components a mass have been studied in \cite{amit};
they also follow from a generalization of the argument for the case $M=1$ given in 
\cite{nelsonpelcovits}. For generic $N$ and $M$ we obtain:
\begin{subequations}\label{RG-flow-equations}
\begin{align}
\frac{dT(\ell )}{d\ell }& =\frac{T(\ell )^{2}}{2\pi }\left( N-M-2+\frac{M}{1+\Delta (\ell )}\right), \label{Tflow}  \\
\frac{d\Delta (\ell )}{d\ell }& =2\Delta (\ell )-\frac{1}{\pi }\frac{T(\ell)
\Delta (\ell )}{1+\Delta (\ell )},
\end{align}
\end{subequations}
where $T(\ell)$ and $\Delta(\ell)$ are the renormalized temperature and anisotropy at length scale $\xi=ae^\ell$ as defined before.
For the present case $N=M+2$, the first flow
equation simplifies to
\begin{equation}
\frac{dT(\ell )}{d\ell }=\frac{T(\ell )^{2}}{2\pi }\frac{N-2}{1+\Delta (\ell)}.\tag{\ref{Tflow}'}
\end{equation}
As long as $\Delta (\ell )$ is small, the flow of the temperature (i.e. of
the inverse stiffness) is that of the usual O($N$) model, while $\frac{dT(\ell )}{d\ell }\rightarrow 0$ at large $\Delta (\ell )$ as
expected for the XY model, where spin wave fluctuations do not renormalize
the stiffness. In this limit, renormalization will only occur via vortex-antivortex fluctuations of
the KT-flow equations, that eventually lead to (\ref{tbkt}).

The solution to the flow equations (\ref{RG-flow-equations}) comes from noticing that $\frac{d}{d\ell }[T(\ell )^{2/M}\Delta (\ell )]=\ 2T(\ell )^{2/M}\Delta (\ell )$,
which allows us to implicitly construct the solution:
\begin{subequations}\label{implicitall}
\begin{gather}
 T(\ell )^{2/M}\Delta (\ell ) e^{-2\ell} = \displaystyle C,  \label{implicit1} \\
 \left( \frac{\left( 1+\Delta (\ell )\right) e^{-\frac{4\pi 
}{MT(\ell )}}}{\Delta (\ell )T(\ell )^{2/M}}+\frac{2E_{1-\frac{2}{M}}\left( 
\frac{4\pi }{MT(\ell )}\right) }{MT(\ell )^{2/M}}\right) =  \displaystyle D,  \label{implicit2}
\end{gather}
\end{subequations}
where $C$ and $D$ are constants determined from the bare parameters when $\ell=0$, and $E_{n}(x)$ is the exponential integral
function.  The RG flow is plotted in Fig.~\ref{Fig1} for $M=2$; other values of $M$ look qualitatively the same.

The RG flow equations (\ref{RG-flow-equations}) are for spin-waves only --- when the anisotropy $\Delta$ reaches a value of order $1$ corresponding to the O(2) phase, this must be supplemented by vortices and BKT arguments.  We therefore stop the flow when $\Delta(\ell_0)=1$ and ask the question:\ what is the renormalized value of temperature $T(\ell_0)$ at this scale?  Above this scale, one sees only the physics of the traditional XY model, and so $T(\ell_0)$ constitutes  the initial condition of the BKT flow \eqref{BKT-RG}.  
As it is not even possible to define a vortex fugacity from the point of view of the $O(N)$ spin waves, the previously discussed fugacity enhancement no longer plays a role.  In fact, from the point of view of the vortices in this approach, $\xi_0=ae^{\ell_0}$ should be considered the microscopic (and only) length scale in the problem.
Hence one finds a usual BKT transition at renormalized temperature $T(\ell_0) \approx \pi/2$.  
Using Eqs.~\ref{implicitall} to trace this RG flow line back to the bare values of temperature and anisotropy, one finds that 
$T_{\mathrm{BKT}}\sim 1/\ln (1/\Delta )$, in agreement with our result based entirely on vortices.

\begin{figure}[tbp]
\begin{center}
\includegraphics[width=2.6in,clip=true]{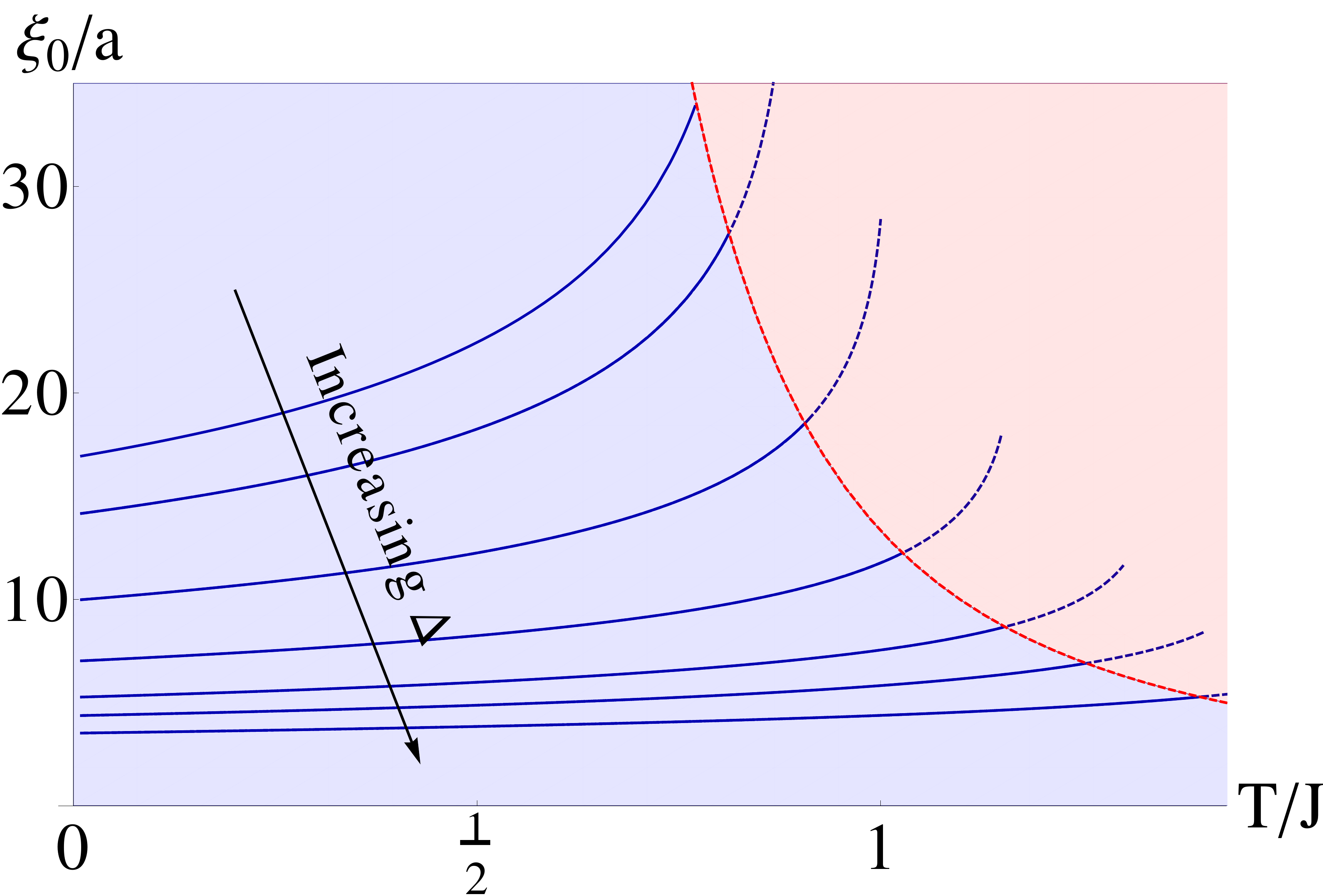}\vspace{-0.6cm}
\end{center}
\caption{Core size $\xi_0$ as a function of temperature $T$ for various values of $\Delta$.  The red dashed line intersects the curves at $T_{\rm BKT}(\Delta)$, showing the size of the vortex cores exactly at the unbinding transition.  The lines end at the point where the vortex core size becomes comparable to the inter-vortex distance, i.e. the point where $\xi$ no longer has any meaning and we are in the crossover from the O(2) to the O($N$) regime.}
\label{Fig3}
\end{figure}

We now discuss this length scale $\xi_0$, which can of course be
associated with the size of the vortex cores. Using (\ref{implicit1}) we have
\begin{equation}
\xi_0^{2}=\frac{a^{2}}{\Delta }\left( \frac{T(\ell _{0})}{T}\right) ^{2/M}.
\end{equation}
Far below the BKT transition temperature, where $T$ barely
renormalizes, we can see that $\xi_0 \simeq a/\sqrt{\Delta }$, again in accordance
with our prior considerations. However, at higher temperatures $T(\ell )$
flows toward strong coupling.   Near $T_{\mathrm{BKT}}$, where $T(\ell
_{0})\simeq 1$ as well, we obtain a logarithmic correction due to spin wave
excitations that immediately leads to our result (\ref{core at BKT}). As
the flow for $\ell <\ell _{0}$ is governed by the RG equation of an O($N$) nonlinear $\sigma$-model, the vortex core enhancement is
dictated by the high symmetry fixed point.
In Fig.~\ref{Fig3} we show the temperature
dependence of the core size.

Above but near $T_{\mathrm{BKT}}$ vortices proliferate, but
the behavior at large distances is still that of an O(2)
problem.  As temperature is further raised, a crossover occurs where
$\Delta(\ell )$ never comes close to being of order unity, even as $\ell
\rightarrow \infty $.  A separatrix (shown in Fig.~\ref{Fig1}) separates regions of flow
where $\Delta (\ell \rightarrow \infty )$ does or does not diverge, indicating the crossover from O(2) to O($N$) behavior \cite{newfootnote}. Analysing this behavior for small $\Delta $ yields that the
crossover temperature also vanishes as $1/\ln (1/\Delta )$, yet with a
larger numerical coefficient:\ see Fig.~\ref{Fig1}. From the perspective of
proliferated vortices, this crossover essentially corresponds to reaching
the regime where the typical intervortex distance is of the order of the
vortex core size, i.e.\ O($N$) fluctuations of the core govern the
entire system. Following the BKT flow equation for the vortex
fugacity confirms this interpretation.

The full agreement between our two approaches is the most important conceptual conclusion of this work -- above the length scale $\xi_0$, the parameters flow with the usual BKT equations \eqref{BKT-RG}; however below this length scale one can either choose to look at the problem from the point of view of vortices, \eqref{simple-RG}, or from spin waves, \eqref{RG-flow-equations}.  In other words, it patches together the perturbative (spin-wave) and non-perturbative (vortex) aspects of the theory, something which has also recently been studied in a completely different context \cite{koenigetal}.

Finally we comment on a peculiarity of the case where $M=2$, i.e.\ where we have
two competing O(2) order parameters. It has been established
that this model at $T=0$ is governed by a tetracritical point, implying
that both order parameters are non-zero at zero temperature for some range of $\Delta$ \cite{fishernelson,kosterlitznelson,Calabrese-2003}. In this case we
have two distinct BKT transitions for the two order parameters, where
the upper transition takes place for the components that are stabilized by
the anisotropy term $\Delta$.  A discussion of this
special case will be given elsewhere \cite{ournext}.

In conclusion, we have analyzed a model of competing order parameters,
at least one of which is of XY type. The energy balance between the competing states
is controlled by a parameter $\Delta$, such that the order parameter
symmetry is enhanced at $\Delta =0$. As the ground state of the system is
fully ordered, we have a situation where the ordering
temperature vanishes as $\Delta \rightarrow 0$ \textit{without} having quantum
critical fluctuations: this vanishing occurs solely because of
the sensitivity with respect to spin waves of distinct order
parameter symmetries. More specifically, the  Berezinskii-Kosterlitz-Thouless transition
temperature $T_{\mathrm{BKT}}$ of vortex proliferation vanishes as $\Delta
\rightarrow 0$ via a divergent core size, i.e. the integrity of
topologically stable vortex configurations is undermined from within. Inside
the giant core of such vortices, high symmetry spin-wave fluctuations
further increase the core size. 

We believe that the observation of giant vortices and of
intra-core excitations (see e.g.~\cite{Lake}) could be an important clue in revealing the competing
nature of order parameters in correlated many body systems.

We would like to thank Peter Orth for useful discussions.
CAH gratefully acknowledges financial support from the EPSRC (UK) via grants EP/I031014/1 and EP/H049584/1.


\begin{thebibliography}{99}
\bibitem{newfootnote}
Using instead the criterion
whether $\Delta (\ell \rightarrow \infty )$ does or does not become unity
only changes numerical coefficients but not the qualitative behavior of this
separatrix.
\end{thebibliography}
\end{document}